\documentclass[10pt]{article}

\usepackage{latexsym}
\usepackage{theorem}
\usepackage{color,graphicx}
\usepackage{amssymb}
\usepackage{amsmath}

\usepackage{txfonts}

\newenvironment{proof}[1][Proof]
{\par\noindent{\bf #1:} }{\hspace*{\fill}\nolinebreak{$\Box$}\bigskip\par}

\newtheorem{theorem}{Theorem}
\newtheorem{lemma}{Lemma}

\theorembodyfont{\upshape}

\begin{document}

\title{The Complexity of Node Blocking for Dags}

\author{Dariusz Dereniowski\\
       Department of Algorithms and System Modeling,\\
       Gda\'{n}sk University of Technology, Poland\\
       \small{deren@eti.pg.gda.pl}}

\maketitle

\begin{center}
\parbox[c]{10 cm}{
\textbf{Abstract:}
We consider the following modification of annihilation game called
node blocking. Given a directed graph, each vertex can be occupied by
at most one token. There are two types of tokens, each player can move
his type of tokens. The players alternate their moves and the current
player $i$ selects one token of type $i$ and moves the token along a
directed edge to an unoccupied vertex. If a player cannot make a move
then he loses. We consider the problem of determining the complexity
of the game: given an arbitrary configuration of tokens in a directed
acyclic graph, does the current player has a winning strategy? We
prove that the problem is PSPACE-complete.
}
\end{center}

\vspace{5 pt}


\section{Introduction}

The study of annihilation games has been suggested by John Conway and
the first papers were published by Fraenkel and Yesha
\cite{FY76,FY82}. They considered a $2$-player game played on an
underlying directed graph $G$ (possibly with cycles). The current
player selects a token and moves it along an arc outgoing from a
vertex containing the token. If a vertex contains two tokens then they
are removed from $G$ (\emph{annihilation}). Authors in \cite{FY82}
gave a polynomial-time algorithm for computing a winning strategy.
In this paper, including all the mentioned here results,
we assume the normal play, where the first player unable
to make a move loses (mis\`ere annihilation games have been considered
in \cite{Ferguson84}).

Fraenkel considered in \cite{Fraenkel02} a generalization of
cellular-automata games to two-player games and provided a strategy
for such cases. In particular, if for each vertex there is at most one
outgoing arc then it is possible to derive a polynomial-time strategy
\cite{Fraenkel02}. Since the formulation of the game is equivalent
to the one mentioned above, this result can be directly applied for
the annihilation game.

Fraenkel in \cite{Fraenkel96} studied the connections between
annihilation games and error-correcting codes. The authors in
\cite{FR03} gave an algorithm for computing error-correcting
codes. The algorithm is polynomial in the size of the code and uses
the theory of two-player cellular-automata games.

In the following we are interested in generalizations of the
annihilation game, where there is more than one type of token and/or
there is a different interaction between the tokens. Assume that
$r\geq 2$ types of tokens are given and each type of token can be
moved along a subset of the edges. Given a configuration of tokens in
a graph, deciding whether the current player has a winning strategy is
PSPACE-complete for acyclic graphs \cite{FG87}.

A modification called \emph{hit}, where $r\geq 2$ types of
tokens and edges are distinguished was considered in \cite{FG87}. A
move consists of selecting a token of type $i$ and moving along an arc
of type $i\in\{1,\ldots,r\}$. The target vertex $v$ cannot be occupied
by a token of type $i$, but if $v$ contains token of other type then
it is removed (so, when the move ends $v$ is occupied by the token of
type $i$). The complexity of determining the outcome of this game is
PSPACE-complete for acyclic graphs and $r=2$ \cite{FG87}.
A modification of hit called \emph{capture} has the same rules except
that each token can travel along any edge. Capture is PSPACE-complete
for acyclic and EXPTIME-complete for general graphs \cite{GR95}.

In a \emph{node blocking} each token is of one of the two
types. Each vertex can contain at most one token. Player $i$ can move
the tokens of type $i$, $i=1,2$. All tokens can move along all
arcs. A player $i$ makes a move, by selecting one token of type $i$
(occupying a vertex $v\in V$) and an unoccupied vertex $u\in V$ such
that $(v,u)\in E$ and moving the token from $v$ to $u$. The
first player unable to make a move loses and his opponent wins the
game. There is a tie if there is no last move. First, the game was
proved to be NP-hard \cite{FY79}, then PSPACE-hard for general
graphs \cite{FG87}. The complexity for general graphs has been finally
proved in \cite{GR95} to be EXPTIME-complete.

In an \emph{edge blocking} all tokens are identical, i.e. each player
can move any token, while each arc is of type $1$ or $2$ and a player
$i$ makes his move by moving a token along an arc of type $i$,
$i=1,2$. Similarly as before, the first player who cannot make a move
loses. A tie occurs if there is no last move. This game is
PSPACE-complete for dags.

The following table summarizes the complexity of all the mentioned
two-player annihilation games. We list only the strongest known
results.
\begin{center}
\begin{tabular}{r||c|c}
Game:         & dag                              & general                   \\\hline\hline
Annihilation  & PSPACE-complete \cite{FG87}      & $?^*$   \\\hline
Hit           & PSPACE-complete \cite{FG87}      & $?^*$   \\\hline
Capture       & PSPACE-complete \cite{GR95}      & EXPTIME-complete \cite{GR95}   \\\hline
Node blocking &       ?                          & EXPTIME-complete \cite{GR95}   \\\hline
Edge blocking & PSPACE-complete \cite{FG87}      & $?^*$   \\\hline
\end{tabular}
\end{center}
Note that for the entries labeled as ``$?^*$'' can be replaced by
``PSPACE-hard'' (which can be concluded from the corresponding results
for acyclic graphs), but the question remains whether the games are in
PSPACE. In this paper we are interested in the problem marked by
``?'', listed also in \cite{playing_games} as one of the open
problems. In Section \ref{sec:node_blocking_hard} we prove
PSPACE-completeness of this game.

\section{Definitions}

In the following a token of type $1$ (respectively $2$) will be called
a \emph{white token} (\emph{black token}, resp.) and denoted by symbol
$W_t$ ($B_t$, resp.). The player moving the white (black) tokens
will be denoted by $W$ ($B$, respectively).

Let $G=(V(G),E(G))$ be a directed graph. For $v\in V(G)$ define
$\deg_G^+(v)=|\{u\in V(G):(u,v)\in E(G)\}|$,
$\deg_G^-(v)=|\{u\in V(G):(v,u)\in E(G)\}|$.
A notation $u\to_p v$ is used to denote a move made by player,
$p\in\{W,B\}$, in which the token has been removed from $u$ and
placed at the vertex $v$. Given the positions of tokens, define
$f(v)$ for $v\in V(G)$ to be one of three possible values
$W_t,B_t,\emptyset$ indicating that a white or black token is at the
vertex $v$ or there is no token at $v$, respectively. In the latter
case we say that $v$ is \emph{empty}. Note that if $f(u)=\emptyset$
or $f(v)\neq\emptyset$ then the move $u\to_p v$ is incorrect.

Let us recall a PSPACE-complete Quantified Boolean Formula
(\emph{QBF}) problem \cite{QBF}. The input for the problem is a
formula $Q$ in the form
\[Q_1x_1 \ldots Q_nx_n F(x_1,\ldots,x_n),\]
where $Q_i\in\{\exists,\forall\}$ for $i=1,\ldots,n$.
Decide whether $Q$ is true.
In our case we us a restricted case of this problem where
$Q_1=\exists$, $Q_{i+1}\neq Q_i$ for $i=1,\ldots,n-1$, $n$
is even, and $F$ is a 3CNF formula, i.e.
$F=F_1\land F_2\land\cdots\land F_m$, where
$F_i=(l_{i,1}\lor l_{i,2}\lor l_{i,3})$ and each literal
$l_{i,j}$ is a variable or the negation of a variable,
$i=1,\ldots,m,j=1,2,3$.

\section{PSPACE-hardness of node blocking}
\label{sec:node_blocking_hard}

\begin{figure}[htb]
\begin{center}
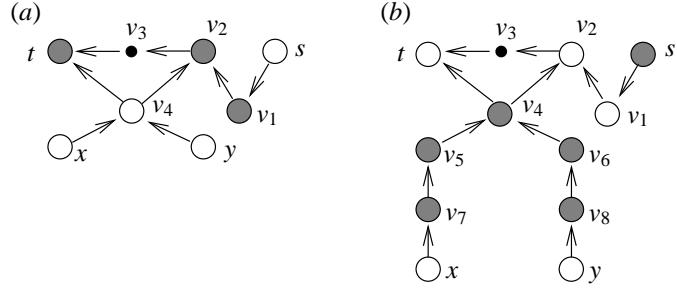
\caption{The graphs $G_i$ for ($a$) $i=2j-1$ and ($b$) $i=2j$, $j=1,\ldots,n/2$}
\label{pic:play_g_i}
\end{center}
\end{figure}

Define a \emph{variable component} $G_i$ corresponding to $x_i$ as
follows:
\[V(G_i)=\{s,t,x,y\}\cup\{v_1,\ldots,v_4\},\]
\[E(G_i)=\{(s,v_1),(v_1,v_2),(v_2,v_3),(v_3,t),(v_4,t),(v_4,v_2),(x,v_4),(y,v_4)\}\]
for $i=2j-1$, and
\[V(G_i)=\{s,t,x,y\}\cup\{v_1,\ldots,v_8\},\]
\begin{eqnarray}
E(G_i) & = & \{(s,v_1),(v_1,v_2),(v_2,v_3),(v_3,t),(v_4,t),(v_4,v_2), \nonumber\\
       &   & (v_5,v_4),(v_6,v_4),(v_7,v_5),(v_8,v_6),(x,v_7),(y,v_8)\} \nonumber
\end{eqnarray}
for $i=2j$, where $j=1,\ldots,n/2$.
Fig.~\ref{pic:play_g_i} depicts these subgraphs.
If $i$ is odd then $G_i$ is called a \emph{white component} and in
this case an initial placement of tokens in $G_i$ is
$f(s)=f(v_4)=f(x)=f(y)=W_t$, $f(v_3)=\emptyset$
and $f(v_1)=f(v_2)=f(t)=B_t$ (see also Fig.~\ref{pic:play_g_i}($a$)).
In a \emph{black component} $G_i$, where $i$ is even, we have
$f(s)=f(v_4)=\ldots=f(v_8)=B_t$, $f(v_3)=\emptyset$
and $f(v_1)=f(v_2)=f(x)=f(y)=f(t)=W_t$
(see also Fig.~\ref{pic:play_g_i}($b$)). In both cases the above
configuration of tokens will be called the \emph{initial state} of
$G_i$.

Removing a token from a graph without placing it on another vertex
is an invalid operation. However, assume for now that, given an initial
state of $G_i$, the first move is a deletion of a token
occupying the vertex $t$ (we will assume in Lemma
\ref{lem:winning_in_component} that the game starts in this
way). Then, $W$ (respectively $B$) becomes the current player in the
white (black, resp.) component $G_i$. Furthermore, we assume that
the game in $G_i$ ends when $f(s)$ becomes $\emptyset$.

\begin{lemma}
If $G_i$ is a white $($respectively black$)$ component then $W$ $(B$,
resp.$)$ has a winning strategy. At the end of the game we have
that if $G_i$ is a white component then exactly one of the vertices
$x,y$ is empty, and if $G_i$ is a black component then
exactly one of the vertices $x,y,v_5,v_6$ is empty.
\label{lem:winning_in_component}
\end{lemma}
\begin{proof}
First assume that $G_i$ is a white component. Let $f(t)=\emptyset$
and $W$ is the current player. The first two moves are
$v_4\to_W t$, $v_2\to_B v_3$. Then there are two possibilities:
\begin{equation}
x\to_W v_4 \textup{ or } y\to_W v_4.
\label{eq:Wbinary_choice}
\end{equation}
In both cases the game continues as follows:
$v_1\to_B v_2$, $s\to_W v_1$. The thesis follows.

Let $G_i$ be a black component with $f(t)=\emptyset$ and $B$ is the
current player. Similarly as before we have $v_4\to_B t$,
$v_2\to_W v_3$. The third move is $v_5\to_B v_4$ or $v_6\to_B
v_4$. Since they are symmetrical, assume in the following that the
first case occurred. We have $v_1\to_W v_2$. Then $B$ has a choice:
\begin{equation}
v_7\to_B v_5\textup{ or }s\to_B v_2.
\label{eq:Bchoice}
\end{equation}
If the first move occurred then we have $x\to_W v_7$. Then,
$s\to_B v_2$, which ends the game and the vertex $x$ is empty
among the vertices listed in the lemma. If $B$ selected the
second move in (\ref{eq:Bchoice}) then the game ends with
$f(v_5)=\emptyset$.
\end{proof}

\begin{figure}[htb]
\begin{center}
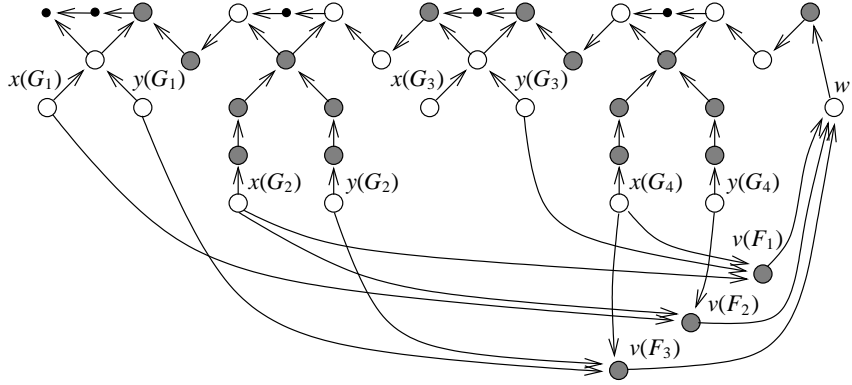
\caption{A complete instance of the graph $G_F$ corresponding to (\ref{eq:Q_ex})}
\label{pic:red_e}
\end{center}
\end{figure}

Now we define a graph $G_F$, corresponding to the Boolean formula $F$.
In order to distinguish a vertex $v\in V(G_i)$ from the vertices of the
other variable components we will write $v(G_i)$.
$G_F$ contains disjoint white components $G_{2i-1}$ for $i=1,\ldots,n/2$
and disjoint black components $G_{2i}$, $i=1,\ldots,n/2$,
connected in such a way that $s(G_i)=t(G_{i+1})$ for $i=1,\ldots,n-1$.
The graph $G_F$ contains additionally the vertices $w,v(F_1),\ldots,v(F_m)$,
an arc $(w,t(G_n))$, the arcs $(v(F_i),w)$ for $i=1,\ldots,m$, and
$(x(G_i),v(F_j))\in E(G_F)$ iff $F_j$ contains $x_i$, while
$(y(G_i),v(F_j))\in E(G_F)$ iff $F_j$ contains $\overline{x_i}$,
a negation of the variable $x_i$.
Initially, all the subgraphs $G_i$ are in the initial state, except
that $f(t(G_1))=\emptyset$. Let $f(w)=W_t$, $f(v(F_j))=B_t$ for
$j=1,\ldots,m$. Before we prove the main theorem, let us demonstrate
the above reduction by giving an example
\begin{equation}
  Q=\exists_{x_1}\forall_{x_2}\exists_{x_3}\forall_{x_4} 
    (x_2\lor\overline{x_3}\lor x_4)\land
    (x_1\lor x_2\lor\overline{x_4})\land
    (\overline{x_1}\lor\overline{x_2}\lor x_4).
\label{eq:Q_ex}
\end{equation}
Fig.~\ref{pic:red_e} shows the corresponding graph $G_F$.

For brevity we introduce a notation: we say that the game
\emph{arrives at} a component $G_i$ (and \emph{leaves}
the component $G_{i-1}$, $i>1$) if $f(t(G_i))=\emptyset$
(note that for $i>1$ this is equivalent to
$f(s(G_{i-1}))=\emptyset$ in the graph $G_F$). The game
\emph{is in} $G_i$ if it arrived at $G_i$ but did not
leave $G_i$.

\begin{theorem}
Node blocking is \textup{PSPACE}-complete for directed acyclic
graphs.
\label{thm:bocking_pspace}
\end{theorem}
\begin{proof}
First we prove by an induction on $i=1,\ldots,n$ that we
may without loss of generality assume that if the game arrives at
the component $G_i$ then
\begin{description}
\item{(i)} for each $j<i$ exactly one of the vertices
           $x(G_j),y(G_j)$ (if $G_j$ is a white component) or
           exactly one of the vertices
           $x(G_j),y(G_j),v_5(G_j),v_6(G_j)$ (if $G_j$ is a black
           component) is empty,
\item{(ii)} all tokens in components $G_j$, for $j=i,\ldots,n$
            are in the initial state, except that $f(t(G_i))=\emptyset$.
\end{description}
The cases for $i=1$ and $i>1$ are analogous. If the game is in $G_i$
then (by the induction hypothesis) all possible moves are the ones
along the arcs in $G_i$, $v_2(G_j)\to_p v_3(G_j)$
for $j>i$ and $v_7(G_j)\to_B v_5(G_j)$ or $v_8(G_j)\to_B v_6(G_j)$ for
a black component $G_j$, $j<i$. In the latter case $W$ responds
$x(G_j)\to_W v_7(G_j)$ or $y(G_j)\to_W v_8(G_j)$, respectively, so we
consider the first two cases.
Let $G_j$ be a white component (the other case is
analogous) and $B$ moves a token along an arc which does not
belong to $E(G_i)$, i.e.
\begin{equation}
v_2(G_j)\to_B v_3(G_j), j>i.
\label{eq:wrong_move}
\end{equation}
For each move (\ref{eq:wrong_move}) $W$ responds
\begin{equation}
v_4(G_j)\to_W v_2(G_j).
\label{eq:wrong_move_response}
\end{equation}
For other moves of $B$, $W$ responds as in the proof of Lemma
\ref{lem:winning_in_component}. Consider the case when the game
arrives at the component which is not in the initial state, because
the moves (\ref{eq:wrong_move}) and (\ref{eq:wrong_move_response})
have been performed. This situation is given in
Fig.~\ref{pic:wrong_move}($a$).
\begin{figure}[htb]
\begin{center}
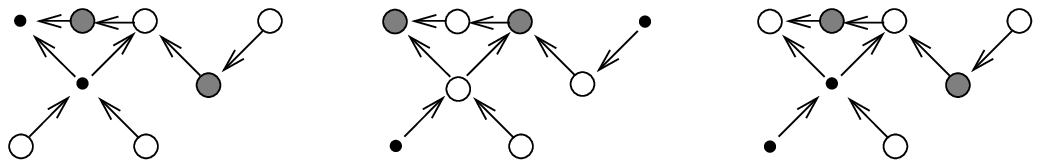
\caption{($a$) the game arrives at $G_j$,
         ($b$) the game leaves $G_j$,
         ($c$) $W$ wins the game}
\label{pic:wrong_move}
\end{center}
\end{figure}
Since $W$ is the current player, the first move in $G_j$ is
$x(G_j)\to_W v_4(G_j)$ or $y(G_j)\to_W v_4(G_j)$. In both cases the
remaining sequence of moves is identical:
$v_3(G_j)\to_B t(G_j)$,
$v_2(G_j)\to_W v_3(G_j)$,
$v_1(G_j)\to_B v_2(G_j)$,
$s(G_j)\to_W v_1(G_j)$.
The result is shown in Fig.~\ref{pic:wrong_move}($b$).
This proves that if $B$ performs a move along an arc which is not in
$G_i$ when the game is in $G_i$ then $W$ decides among one of the
moves $x(G_j)\to_W v_4(G_j)$ or $y(G_j)\to_W v_4(G_j)$ when the game
is in $G_j$. This, however is only true under the assumption that
after (\ref{eq:wrong_move}) and (\ref{eq:wrong_move_response}) $W$
plays according to the schema given in the proof of Lemma
\ref{lem:winning_in_component}. If the white player managed to place
a token at the vertex $v_4(G_j)$ before the game arrived at $G_j$
then the move $v_4(G_j)\to_W t(G_j)$ gives a situation depicted in
Fig.~\ref{pic:wrong_move}($c$) --- the black player cannot make a move
in $G_j$. So, if the game is in $G_i$ and a move
(\ref{eq:wrong_move}) occurred, then either the game creates the same
configuration of tokens in variable components (restricted to the
vertices $x(G_k),y(G_k),k=1,\ldots,n$), or $B$ loses the game. Thus,
w.l.o.g. we may assume that if the game is in $G_i$ then the
components $G_j$, $j>i$ are in the initial state, i.e. (ii) is true.

Assuming the players make only moves along the arcs of
$G_i$, if the game arrives at $G_{i+1}$ then Lemma
\ref{lem:winning_in_component} implies that (i) is
satisfied.

Now we can prove the theorem. Assume that $Q$ is true and we show that
$W$ has a winning strategy. If $x_i$ is true (respectively false),
$i=2k-1$, $k=1,\ldots,n/2$, then $W$ plays in $G_i$ in such a way that
if the game leaves $G_i$ then $f(x(G_i))=W_t$ ($f(y(G_i))=W_t$,
respectively). Assume that the game
leaves $G_n$. Then we have $w\to_W s(G_n)$ and $v(F_j)\to_B w$, for some
$j\in\{1,\ldots,m\}$. Since $Q$ is true, there is a true literal $l_{j,k}$
in $F_j$, $k\in\{1,2,3\}$. If $l_{j,k}=x_t$ then $f(x(G_t))=W_t$ and
$W$ can make the move $x(G_t)\to_W v(F_j)$. If $l_{j,k}=\overline{x_t}$
then $f(y(G_t))=W_t$ and the move $y(G_t)\to_W v(F_j)$ is possible.
Note that if $x(G_t)$ or $y(G_t)$ belongs to a black component, then
(because $Q$ is true) $W$ always has a possibility to
make the above move in such a way that it holds
$f(v_5(G_t))=B_t$ or $f(v_6(G_t))=B_t$, respectively. If $B$ can make a
move then it must be $v_7(G_j)\to_B v_5(G_j)$ or $v_8(G_j)\to_B v_6(G_j)$,
but then $W$ responds $x(G_j)\to_B v_7(G_j)$ or $y(G_j)\to_B v_8(G_j)$.
No other moves are possible, so $W$ wins the game. The above holds for
each index $j$.

Let now $W$ have a winning strategy. If the values of
$x_1,\ldots,x_i$, $i=2k$ have been set then let
$x_{i+1}=\textup{true}$ if we have the move $y(G_{i+1})\to_W
v_4(G_{i+1})$ during the game in $G_{i+1}$, and let
$x_{i+1}=\textup{false}$ if there is a move
$x(G_{i+1})\to_W v_4(G_{i+1})$  during the game in $G_{i+1}$.
The game leaves $G_n$ and we have the moves $w\to_W s(G_n)$,
$v(F_j)\to_W w$ for some $j\in\{1,\ldots,m\}$. The black player chooses
$j$ arbitrarily and since $W$ has a winning strategy there is possible
a move $x(G_k)\to_W v(F_j)$ or $y(G_k)\to_W v(F_j)$. From the construction
of the strategy for $W$ we have that there is the literal
$x_k=\textup{true}$ in $F_j$ or the literal
$\overline{x_k}=\textup{true}$ in $F_j$, respectively.

Observe that $|V(G_F)|=7n/2+11n/2+m+2$, so this is a polynomial
reduction. This proves PSPACE-hardness of node blocking. One can
argument that $G_F$ is acyclic which implies that the game is in
PSPACE.
\end{proof}

\bibliographystyle{plain}
\bibliography{games}

\end{document}